\documentclass[nofootinbib,twocolumn,amsmath,showpacs]{revtex4-2}
\usepackage{graphicx}
\usepackage{dcolumn}
\usepackage{bm}
\usepackage{color} 
\usepackage{slashed}
\begin{document}
\newcommand{\hs}{\hspace*{0.5cm}}
\newcommand{\vs}{\vspace*{0.5cm}}
\newcommand{\be}{\begin{equation}}
\newcommand{\ee}{\end{equation}}
\newcommand{\bea}{\begin{eqnarray}}
\newcommand{\eea}{\end{eqnarray}}
\newcommand{\ben}{\begin{enumerate}}
\newcommand{\een}{\end{enumerate}}
\newcommand{\bde}{\begin{widetext}}
\newcommand{\ede}{\end{widetext}}
\newcommand{\nn}{\nonumber}
\newcommand{\crn}{\nonumber \\}
\newcommand{\Tr}{\mathrm{Tr}}
\newcommand{\non}{\nonumber}
\newcommand{\noi}{\noindent}
\newcommand{\al}{\alpha}
\newcommand{\la}{\lambda}
\newcommand{\bet}{\beta}
\newcommand{\ga}{\gamma}
\newcommand{\va}{\varphi}
\newcommand{\om}{\omega}
\newcommand{\pa}{\partial}
\newcommand{\+}{\dagger}
\newcommand{\fr}{\frac}
\newcommand{\bc}{\begin{center}}
\newcommand{\ec}{\end{center}}
\newcommand{\Ga}{\Gamma}
\newcommand{\de}{\delta}
\newcommand{\De}{\Delta}
\newcommand{\ep}{\epsilon}
\newcommand{\varep}{\varepsilon}
\newcommand{\ka}{\kappa}
\newcommand{\La}{\Lambda}
\newcommand{\si}{\sigma}
\newcommand{\Si}{\Sigma}
\newcommand{\ta}{\tau}
\newcommand{\up}{\upsilon}
\newcommand{\Up}{\Upsilon}
\newcommand{\ze}{\zeta}
\newcommand{\ps}{\psi}
\newcommand{\Ps}{\Psi}
\newcommand{\ph}{\phi}
\newcommand{\vph}{\varphi}
\newcommand{\Ph}{\Phi}
\newcommand{\Om}{\Omega}
\newcommand{\AdrHEPC}{Phenikaa Institute for Advanced Study, Phenikaa University, Nguyen Trac, Duong Noi, Hanoi 100000, Vietnam}

\title{Twin hypercharges} 

\author{Nguyen Huy Thao}
\email{nguyenhuythao@hpu2.edu.vn}
\affiliation{Faculty of Physics, Hanoi Pedagogical University 2, Xuan Hoa, Phu Tho, Vietnam}

\author{Nguyen Thi Nguyet Nga}
\email{nganguyet4988@hvu.edu.vn}
\affiliation{Faculty of Natural Sciences, Hung Vuong University, Nong Trang, Phu Tho, Vietnam}

\author{Tran Dinh Tham}
\email{tdtham@pdu.edu.vn}
\affiliation{Faculty of Natural Science Teachers' Education,
Pham Van Dong University, 509 Phan Dinh Phung,
Cam Thanh, Quang Ngai, Vietnam} 

\author{Phung Van Dong} 
\email{dong.phungvan@phenikaa-uni.edu.vn (corresponding author)}
\affiliation{\AdrHEPC} 

\date{\today}

\begin{abstract}

It is shown that a duplication of the hypercharge, which is identical for the normal sector but different for the dark sector, may manifestly address neutrino mass and dark matter.  

\end{abstract} 

\maketitle

The presence of the hypercharge is only the way consistently combining the weak isospin and the electric charge performing the electroweak symmetry, as well as protecting the unitarity at high energy, which directly predicts the existence of the weak neutral current, leading to the success of the standard model \cite{glashow,weinberg,salam}. However, this proposal also leaves profound questions of the physics unanswered. As the electric charge, the hypercharge is abelian, thus not fixed. The hypercharge was indeed chosen to describe the discrete values of electric charge, as observed, while it cannot explain them, the problem of charge quantization. Further, it predicts massless neutrinos, which oppose the experiments of neutrino oscillations \cite{kajita,mcdonald}. There is no any candidate for dark matter, which makes up most of the mass of our universe \cite{bertone,arcadi}. 

The symmetry of weak isospin $T_{1,2,3}$ motivated extending the V$-$A theory \cite{fg,sm,sakurai} demands that left-handed fermions $l_L=(\nu_L,e_L)$ and $q_L=(u_L,d_L)$ transform as its doublets, while it puts relevant right-handed fermions in its singlets. Hence, the electric charge of the doublets is obtained as $Q=\mathrm{diag}(0,-1)$ for $l_L$ and $Q=\mathrm{diag}(2/3,-1/3)$ for $q_L$. Hence, $[Q,T_1\pm i T_2]=\pm (T_1\pm i T_2)\neq 0$ and $\mathrm{Tr}Q\neq 0$, which yield that the electric charge neither commutes nor closes algebraically with the weak isospin, respectively. Further, it is verified that $[Q-T_3,T_1\pm i T_2]=0$, implying the existence of a new abelian charge, $Y=Q-T_3$. Although this minimal choice of $Y$ successfully predicts the weak neutral current, it is not the end of the story. As the electric charge, $Y$ is not fixed due to the nature of abelian algebra, $[Y,Y]=0$. Hence, a curious question arising is that can the mentioned new physics come from the presence of multiple of the hypercharge, instead of just the one as in the usual theory? To realize this hypothesis, we suggest that there is a duplication of hypercharge, called $Y_{1}$, $Y_{2}$, which must act the same for normal sector but distinct for dark sector. $Y_{1,2}$ are spontaneously broken down to the usual hypercharge $Y=(Y_1+Y_2)/2$ beside a residual dark parity $P_D=(-1)^D$ for $D=(Y_1-Y_2)/2$. It is clear that the normal sector is parity even, while the dark sector is parity odd. This both generates radiative neutrino masses and provides dark mater candidates. 

Since the hypercharge has a nature different from gauged lepton/baryon charges, such as $B-L$, $L_i-L_j$, and $\fr 1 3 (B_i-B_j)$, the proposal of the twin hypercharges is distinct from the latter charges. Furthermore, the difference of the twin hypercharges results in a dark charge, which provides a novel origin of the well-studied dark charge \cite{dc1,dc2,dc3}. It is interesting that this proposal may arise from a more fundamental theory, such as little Higgs model where multiple copies of the standard model gauge group are imposed \cite{lh0,lh1,lh2}. It is noteworthy that the dark charge has a nature distinct from $T$-parity in the little Higgs \cite{lh3,lh4,lh5}, since $T$-parity exchanges repeated gauge groups, whereas the dark parity is a residual gauge symmetry defined by a new nontrivial vacuum which exists without necessarily imposing such exchange symmetry.    

Including the color group, the full gauge symmetry takes the form, 
\be SU(3)_C\otimes SU(2)_L\otimes U(1)_{Y_1}\otimes U(1)_{Y_2}.\label{fgs}\ee The hypercharges determine the electric charge, such as $Q=T_3+\fr 1 2 (Y_1+Y_2)$, where $Y_{1,2}$ are identical to that of the standard model. However, for the new sector, we impose a vectorlike fermion $N_{L,R}$ for each family, possessing opposite $Y_{1,2}$, which make the electric charge of $N_{L,R}$ to be vanished, as expected. Two scalar doublets are included, one acts as in the Higgs mechanism, while the other one that couples $\nu_L$ to $N_R$ induces neutrino masses, where the two doublets necessarily couple to a scalar singlet $\xi$. Additionally, $Y_{1,2}$ are broken by $\chi$ that couples to $N$'s by themselves. That said, the particle representations under the full gauge symmetry are collected in Table \ref{tab1}. Here, the dark parity $P_D=(-1)^{(Y_1-Y_2)/2}$ is also included. 
\begin{table}[h]
\begin{tabular}{cccccc}
\hline\hline
Field & $SU(3)_C$ & $SU(2)_L$ & $U(1)_{Y_1}$ & $U(1)_{Y_2}$ & $P_D$\\
\hline
$l_L=(\nu_L,e_L)$ & 1 & 2 & $-1/2$ & $-1/2$ & $+$\\
$q_L=(u_L,d_L)$ & 3 & 2 & $1/6$ & $1/6$ & $+$\\
$e_R$ & 1 & 1 & $-1$ & $-1$ & $+$\\
$u_R$ & 3 & 1 & $2/3$ & $2/3$ & $+$\\
$d_R$ & 3 & 1 & $-1/3$ & $-1/3$ & $+$\\
$N$ & 1 & 1 & $-1$ & $+1$ & $-$\\
$\phi=(\phi^+_1,\phi^0_2)$ & 1 & 2 & $1/2$ & $1/2$ & $+$\\
$\eta=(\eta^0_1,\eta^-_2)$ & 1 & 2 & $1/2$ & $-3/2$ & $-$\\
$\xi$ & 1 & 1 & $-1$ & $+1$ & $-$\\
$\chi$ & 1 & 1 & $+2$ & $-2$ & $+$\\
\hline\hline
\end{tabular}
\caption[]{\label{tab1} Particle representation content of the model.}
\end{table}         

The scalar potential takes the form,
 \bea V &=& \mu^2_1\phi^\dagger \phi +\mu^2_2 \eta^\dagger \eta +\mu^2_3\chi^*\chi+\mu^2_4\xi^*\xi \crn
&&+ \la_1 (\phi^\dagger \phi)^2+\la_2 (\eta^\dagger \eta)^2+\la_3 (\chi^*\chi)^2+\la_4 (\xi^*\xi)^2\crn
&&+ \la_5 (\phi^\dagger \phi)(\eta^\dagger \eta)+\la_6 (\phi^\dagger \eta)(\eta^\dagger \phi)+\la_7(\phi^\dagger \phi)(\chi^*\chi)\crn
&&+\la_8 (\phi^\dagger \phi)(\xi^*\xi)+\la_9(\eta^\dagger \eta)(\chi^*\chi)+\la_{10}(\eta^\dagger \eta)(\xi^*\xi)\crn
&&+\la_{11}(\chi^* \chi)(\xi^*\xi) + (\kappa_1 \phi \eta \xi +\kappa_2 \xi\xi\chi+H.c.),\eea where $\la$'s are dimensionless, while $\mu$'s and $\kappa$'s have a mass dimension. The condition for the potential to be bounded from below, as well as yielding an expected vacuum structure, requires $\mu^2_{1,3}<0$, $\mu^2_{2,4}>0$, $\la_{1,2,3,4}>0$, and other conditions for $\la$'s such that the scalar quartic coupling matrix is copositive \cite{kannike}. Hence, the scalar fields develop vacuum expectation values (VEVs), such as $\langle \phi\rangle = (0,v/\sqrt{2})$, $\langle \chi\rangle =\La/\sqrt{2}$, $\langle \eta\rangle =0$, and $\langle \xi\rangle =0$, where $v^2=2(\la_7\mu^2_3-2\la_3\mu^2_1)/(4\la_1\la_3-\la^2_7)$ and $\La^2=2(\la_7\mu^2_1-2\la_1\mu^2_3)/(4\la_1\la_3-\la^2_7)$ are obtained by the conditions of potential minimization. To be consistent with the standard model, we impose $\La\gg v$. 

The scheme of symmetry breaking is 
\bc \begin{tabular}{c}
$SU(3)_C\otimes SU(2)_L\otimes U(1)_{Y_1}\otimes U(1)_{Y_2}$\\
$\downarrow\La$\\
$SU(3)_C\otimes SU(2)_L\otimes U(1)_{Y}\otimes P_D$\\
$\downarrow v$\\
$SU(3)_C\otimes U(1)_{Q}\otimes P_D$
\end{tabular} \ec It is noted that $Y=(Y_1+Y_2)/2$ annihilates the vacuum $\La$, since $Y\La=0$, so it is conserved after the first stage of symmetry breaking. Besides this charge, there is a residual symmetry taking the form $P_D=e^{i \om D}$, where $D\equiv (Y_1-Y_2)/2$, and $\om$ is a transformation parameter. $P_D$ conserves $\La$ if $P_D\La=e^{2i\om}\La=\La$, thus $\om=k\pi$ for $k$ integer. Therefore, $P_D=(-1)^{kD}=\{1,(-1)^D\} \cong Z_2$. That said, there is a discrete residual symmetry, besides $Y$, redefined by $P_D=(-1)^{D}$. The usual fields have even dark parity, while the new fields possess odd dark parity, as shown in Table \ref{tab1}. After the second stage of symmetry breaking, the standard model gauge symmetry is broken down to $SU(3)_C\times U(1)_Q$, in which $Q=T_3+Y$, as usual, while $P_D$ is always conserved by $v$. The conservation of $P_D$ ensures that $\eta,\xi$ cannot develop any VEV.  

The covariant derivative is defined by $D_\mu=\pa_\mu+ig_s t_n G_{n\mu} + i g T_a A_{a\mu}+i g_1 Y_1 B_{1\mu}+i g_2 Y_2 B_{2\mu}$, where $(g_s,g,g_1,g_2)$, $(t_n,T_a,Y_1,Y_2)$, and $(G_n,A_a,B_1,B_2)$ are gauge couplings, gauge charges, and gauge bosons according to the gauge subgroups in (\ref{fgs}), respectively. Let $t_1\equiv g_1/g$ and $t_2\equiv g_2/g$. The charged gauge boson is given by $W^\pm=(A_1\mp i A_2)/\sqrt{2}$ with mass $m_W=gv/2$, as usual. Neutral gauge bosons are obtained by 
\bea && A=s_W A_3 + c_W(c_D B_1+s_D B_2),\\
&&Z=c_W A_3 - s_W(c_D B_1+s_D B_2),\\
&& Z'=s_D B_1 - c_D B_2,\eea which correspond to photon, $Z$-boson, and new neutral gauge boson, respectively. Here, the Weinberg angle $\theta_W$ is given by $t_W=2t_1 t_2/\sqrt{t^2_1+t^2_2}$, while the dark angle $\theta_D$ is defined as $t_D=t_1/t_2$. There is a small mixing between $Z$ and $Z'$ through their mass matrix, 
\be M^2_{\mathrm{g}}=\begin{pmatrix}
m^2_Z & m^2_{ZZ'}\\
m^2_{ZZ'} & m^2_{Z'}
\end{pmatrix},\ee where $m^2_Z=g^2 v^2/4c^2_W$, $m^2_{ZZ'}=(g_1 g_2/2)(c_{2D}/s_W)v^2$, and $m^2_{Z'}=[(g^4_1+g^4_2)(16\La^2+v^2)+2g^2_1 g^2_2(16\La^2-v^2)]/4(g^2_1+g^2_2)\simeq 4(g^2_1+g^2_2)\La^2$. This mass matrix is diagonalized to yield physical fields, $Z_1=c_\al Z-s_\al Z'$, $Z_2=s_\al Z+c_\al Z'$, where the mixing angle is given by $t_{2\al}=2m^2_{ZZ'}/(m^2_{Z'}-m^2_{Z})\simeq (gc_{2D}/8c_W\sqrt{g^2_1+g^2_2})(v^2/\La^2)\sim v^2/\La^2$, while the physical masses are $m^2_{Z_{1,2}}=\fr 1 2 [m^2_Z+m^2_{Z'}\mp \sqrt{(m^2_Z-m^2_{Z'})^2+4m^4_{ZZ'}}]$, which yield $m_{Z_1}\simeq gv/2c_W$ and $m_{Z_2}\simeq 2\sqrt{g^2_1+g^2_2}\La$, corresponding to $Z$-like boson and new neutral gauge boson. It is clear that the $Z$-$Z'$ mixing vanishes if $c_{2D}=0$, i.e. $g_1=g_2$. This condition allows $T$-parity working. However, this work has a residual parity even if $g_1\neq g_2$, which is a new observation of this work.               

Expanding around the VEVs, $\phi_2=(v+S+i A)/\sqrt{2}$ and $\chi=(\La +S'+iA')/\sqrt{2}$, the physical parity-even scalar fields are given by   
\bea &&\phi=\begin{pmatrix} G^+_W\\
\fr{1}{\sqrt{2}}(v+c_\beta H+s_\beta H'+ i G_Z)
\end{pmatrix},\\ 
&&\chi=\fr{1}{\sqrt{2}}(\La -s_\beta H+c_\beta H' + i G_{Z'}),\eea where $G^+_W\equiv \phi^+_1$, $G_Z\equiv A$, and $G_{Z'}\equiv A'$ are massless Goldstone bosons according to gauge bosons $W^+$, $Z$, and $Z'$, respectively. $H \equiv c_\beta S - s_\beta S'$ and $H'\equiv s_\beta S +c_\beta S'$ are identified with the usual and new Higgs fields, possessing masses $m^2_{H}\simeq (2\la_1-\la^2_7/2\la_3)v^2$ and $m^2_{H'}\simeq 2\la_3 \La^2$, respectively. The usual and new Higgs mixing angle is determined by $t_{2\beta}\simeq (\la_7 v)/(\la_3 \La) \ll 1$. 

Expanding neutral parity-odd scalars, $\eta_1=(R+i I)/\sqrt{2}$ and $\xi=(R'+i I')/\sqrt{2}$, as well as defining $\mu^2_\eta\equiv \mu^2_2+\fr{\la_5}{2}v^2+\fr{\la_{9}}{2}\La^2$ and $\mu^2_\xi\equiv \mu^2_4+\fr{\la_8}{2}v^2+\fr{\la_{11}}{2}\La^2$, the charged parity-odd scalar $C^-\equiv \eta^-_2$ is itself a physical field with mass $m^2_{C}=\mu^2_\eta +\fr{\la_{6}}{2}v^2$, whereas neutral parity-odd scalars $R,R'$ and $I,I'$ mix in each pair with mixing angles $\theta_R$ and $\theta_I$ defined, respectively, by \be t_{2 R/2I}=\fr{\mp\sqrt{2}\kappa_1 v}{\mu^2_{\xi}\pm\sqrt{2}\kappa_2 \La - \mu^2_\eta},\ee which obey  $\theta_{R,I}\sim v/\La \ll 1$.
The physical neutral parity-odd scalars are $R_1=c_{R} R -s_{R} R'$, $R_2=s_{R} R +c_{R} R'$, $I_1=c_{I} I -s_{I} I'$, and $I_2=s_{I} I +c_{I} I'$, with masses,
\bea && m^2_{R_{1}/I_1}\simeq \mu^2_\eta + \fr{\kappa^2_1 v^2/2}{\mu^2_\eta - \mu^2_\xi\mp\sqrt{2}\kappa_2 \La},\\ 
&& m^2_{R_{2}/I_2}\simeq \mu^2_\xi \pm\sqrt{2} \kappa_2 \La + \fr{\kappa^2_1 v^2/2}{\mu^2_\xi \pm\sqrt{2} \kappa_2 \La - \mu^2_\eta}.\eea 
Notice that $\kappa_{1,2}$ are not prevented by any current symmetry, being as large as the big scale $\kappa_1\sim \kappa_2\sim \La$. However, the mass splitting $(m^2_{R_1}-m^2_{I_1})/(m^2_{R_1}+m^2_{I_1})\sim (v /\La)^2 \ll 1$ is suppressed as $\theta^2_{R,I}\sim (v/\La)^2\ll 1$ is.

On the other hand, the Yukawa couplings are
\bea \mathcal{L}_{\mathrm{Yuk}} &=& h^e \bar{l}_L \phi e_R + h^d \bar{q}_L \phi d_R + h^u \bar{q}_L \tilde{\phi} u_R \crn && + h \bar{l}_L \eta N_R  - M \bar{N}_L N_R +h_L \bar{N}_L\chi^* N^c_L\crn 
 &&+h_R \bar{N}^c_R\chi N_R + H.c.,\eea where $h$'s are dimensionless, while $M$ possesses a mass dimension. The parity-even charged leptons $e$'s and quarks $u$'s, $d$'s gain an appropriate mass similar to the standard model. The couplings $h_{L,R}$ violate lepton number, which would be small. Hence, the corresponding $N_{L,R}$ Majorana masses, labelled $\mu_{L,R}=-\sqrt{2}h_{L,R}\La$, must be radically smaller than $M\sim \La$. The parity-odd fermions $(N^c_L,N_R)$ obtain a mass matrix in such basis as
\be M_{N}=\begin{pmatrix}
\mu_L & M \\
M & \mu_R\end{pmatrix},\ee Since $\mu_{L,R}\ll M$, the fields $N_{L,R}$ act as quasi-Dirac states, related to mass eigenstates, such as $N_{1R}=c_\varphi N^c_L-s_\varphi N_R$, $N_{2R}=s_\varphi N^c_L+c_\varphi N_R$, where the mixing angle is defined by $\cot({2\varphi}) = (\mu_R-\mu_L)/2M \ll 1$, i.e. $\varphi \simeq \fr{\pi}{4}+\fr{\mu_L-\mu_R}{4M}$, or $s_\varphi \simeq c_\varphi \simeq 1/\sqrt{2}$ up to $\mu_{L,R}/M$ order. The physical parity-odd fermions $N_{1,2}$ obtain a mass, approximated as 
\be m_{N_1/N_2}\simeq \mp M+\fr 1 2 (\mu_L+\mu_R),\ee which are opposite at the leading order (i.e., a quasi-Dirac fermion is equivalent to two Majorana states with nearly-opposite masses).

\begin{figure}[h]
\bc
\includegraphics[scale=0.55]{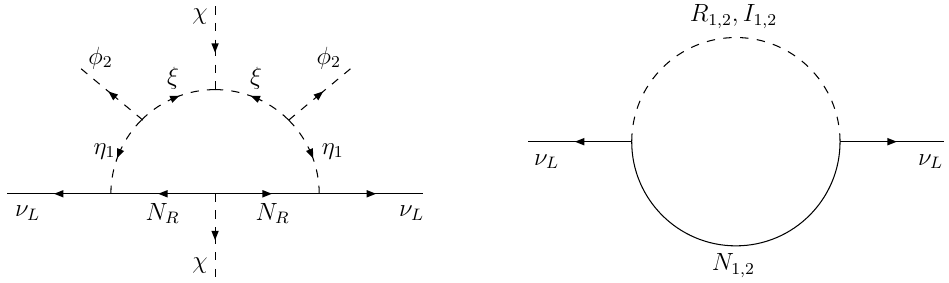}
\caption[]{\label{fig1} Neutrino mass generation scheme in which the left and right diagrams are given in the flavor and mass eigenbases, respectively.}
\ec
\end{figure}
Neutrino mass is generated by a Feynman diagram in Fig. \ref{fig1}. It is evaluated as 
\bea m_\nu &=& \fr{h^2\mu}{32\pi^2}[c^2_R f(M,m_{R_1})-c^2_I f(M,m_{I_1})\crn
&&+s^2_R f(M,m_{R_2})-s^2_I f(M,m_{I_2})],\eea where $\mu\equiv (\mu_L+\mu_R)/2$, and \be f(M,x)=\fr{x^2}{M^2-x^2}\left(2-\fr{M^2+x^2}{M^2-x^2}\ln\fr{M^2}{x^2}\right)\ee is a loop function. Besides the divergences associated with each diagram are manifestly cancelled out by the contributions of real and imaginary parts of scalar fields, the neutrino mass is substantially suppressed by quasi-Dirac fermion fields as contributed by nearly-opposite Majorana masses proportional to $\mu$. The above result is written for a family, but it can be generalized for three families. Since $s^2_R\sim s^2_I\sim (m^2_{R_1}-m^2_{I_1})/(m^2_{R_1}+m^2_{I_1})\sim (v/\La)^2\ll 1$, we approximate $m_\nu\sim \fr{h^2}{32\pi^2}\fr{\mu}{\La}\fr{v^2}{\La}$. Hence, besides the seesaw suppression \cite{cs1,cs2,cs3,cs4,cs5}, the neutrino mass is additionally suppressed by the loop factor $1/16\pi^2$ and the quasi-Dirac approximation $\mu/\La$, which is a new observation of this work, in agreement to \cite{dtn}. For instance, taking $\mu/\La\sim 10^{-4}$, $m_\nu\sim 0.1$ eV requires $h\sim 10^{-2}$, which is sizable, making phenomenological processes viable, opposite to the usual scotogenic setup \cite{sct1,sct2}. 

New prediction of this model is a quasi-Dirac fermion dark matter candidate, called $N_{1}$. That said, $N_{1}$ is the lightest of the dark fields and is stabilized by the dark parity conservation. It dominantly interacts with normal matter via the $Z'$ portal, such as
\bea \mathcal{L} &\supset& -\sqrt{g^2_1+g^2_2} \bar{N}_{1}\ga^\mu N_2 Z'_\mu+H.c.\crn
&&+c_{2D} \sqrt{g^2_1+g^2_2}\bar{f}\ga^\mu\left(Q-\fr 1 2 T_3+\fr 1 2 T_3 \ga_5\right)f Z'_\mu,\eea where $f$ denotes usual quarks and leptons.\footnote{Notice that $N_{1,2}$ are right-handed but their handedness is suppressed for simplicity, while $T_3$ is that for left-handed fermion.} Further, the mixing effect of $Z'$ and $Z$ is small, as suppressed. In the early universe, the co-annihilation of $N_1 N_2$ to normal matter is the most important process, which sets the relic density. It is noted that the annihilations of $N_1 N_1$ and $N_2 N_2$ are strongly suppressed, as they do not directly interact with $Z'$. [Hence, the relevant $s$-channels are $p$-wave suppressed, while the $t$-channels if viable are suppressed by dark matter mass scale and subleading.] It follows that the annihilation cross section is governed by $N_1N_2\to f\bar{f}$, yielding \be \langle \sigma v \rangle\simeq \fr{5 c^2_{2D}(g^2_1+g^2_2)^2m^2_N}{4\pi [(4m^2_{N}-m^2_{Z'})^2+m^2_{Z'}\Ga^2_{Z'}]},\ee where $m_N\simeq m_{N_1}\simeq m_{N_2}\simeq M$. The $Z'$ mass resonance is crucial to set the dark matter relic density, hence $m_{N}\simeq \fr 1 2 m_{Z'}$ is predicted. The dark matter direct detection \cite{bjma} measures the scattering cross section of dark matter with nucleons confined in nuclei, $N_1 \mathcal{N}\to N_2 \mathcal{N}$, for $\mathcal{N}=p,n$, via $Z'$ portal. It is evaluated as $\sigma^{\mathrm{SI}}_{N}\simeq (\sqrt{|g^2_2-g^2_1|}/0.14)^4 (2\ \mathrm{TeV}/m_{Z'})^4\times 10^{-46}\ \mathrm{cm}^2$ (cf. \cite{d331} for an evaluation). The current search implies that a TeV dark matter with a weak coupling may easily evade the bound $\sigma^{\mathrm{SI}}_{\mathrm{exp}}\sim 10^{-46}\ \mathrm{cm}^2$ \cite{lux}. Alternatively, a quasi-Dirac dark matter mass splitting $\Delta m = |m_{N_2}|-|m_{N_1}|=2\mu\sim 10^{-4} \La \gtrsim 100$ MeV for $\La\gtrsim 1$ TeV makes the direct detection cross section kinematically forbidden \cite{msplt}, which is in good agreement with the neutrino mass constraint.

Last, but not least, the precision electroweak test bounds the $\rho$-parameter, which comes from a tree-level mixing between $Z$ and $Z'$, to be $\Delta\rho=\rho-1\simeq m^4_{ZZ'}/(m^2_Z m^2_{Z'})\simeq (c^2_{2D}/16)(v^2/\La^2)\lesssim 0.0004$~\cite{pdg}. It leads to $c_{2D} v/\La\lesssim 0.08$, which is easily satisfied since $|c_{2D}|\leq 1$ and $v/\La\lesssim 0.1$ as appropriately chosen. The $Z$-$Z'$ mixing also modifies the well-measured couplings of $Z$ with fermions. The $Z$-pole measurements limit the corresponding mixing angle $\al\simeq (gc_{2D}/16c_W\sqrt{g_1^2+g_2^2})(v^2/\La^2) \sim 10^{-3}$ \cite{pdg}, which is in agreement to the bound for $\rho$-parameter, given that $g\sim \sqrt{g^2_1+g^2_2}$. The LEPII experiment studies the $Z'$ contribution to process $e^+ e^- \to \mu^+\mu^-$, giving the bound on effective couplings, $\mathcal{L}_{\mathrm{eff}}\supset a_{LL}(\bar{e}\ga^\mu P_L e)(\bar{\mu}\ga_\mu P_L \mu)+(LR)+(RL)+(RR)$, such as $a_{RR}<1/(6\ \mathrm{TeV})^2$ \cite{lepii,carena}. Here note that $4a_{LL}=2a_{LR/RL}=a_{RR}=c^2_{2D}(g^2_1+g^2_2)/m^2_{Z'}$. This is translated to $c_{2D}\sqrt{g^2_1+g^2_2}/m_{Z'}<1/6\ \mathrm{TeV}$, i.e. $\La>c_{2D}\times 3$ TeV, as expected.            

Finally, our understanding of neutrino mass and dark matter might come from the theory of twin hypercharges. The UV-completion of the theory is straightforward for any symmetry that contains the twin hypercharges, in which the dark charge and dark parity automatically results from symmetry breaking.  

This research is funded by Vietnam National Foundation for Science and Technology Development (NAFOSTED) under grant number 103.01-2023.50.

\end{document}